Evaluating the Precision of Estimators of Quantile-Based Risk Measures

By

Kevin Dowd and John Cotter[*]

Abstract

This paper examines the precision of estimators of Quantile-Based Risk Measures (Value at Risk, Expected Shortfall, Spectral Risk Measures). It first addresses the question of how to estimate the precision of these estimators, and proposes a Monte Carlo method that is free of some of the limitations of existing approaches. It then investigates the distribution of risk estimators, and presents simulation results suggesting that the common practice of relying on asymptotic normality results might be unreliable with the sample sizes commonly available to them. Finally, it investigates the relationship between the precision of different risk estimators and the distribution of underlying losses (or returns), and yields a number of useful conclusions.

May, 2007

[*] Kevin Dowd is at the Centre for Risk and Insurance Studies, Nottingham University Business School, Jubilee Campus, Nottingham NG8 1BB, UK; email: Kevin.Dowd@nottingham.ac.uk. John Cotter is at the Centre for Financial Markets, Smurfit School of Business, University College Dublin, Carysfort Avenue, Blackrock, Co. Dublin, Ireland; email: john.cotter@ucd.ie. Dowd's contribution was supported by an Economic and Social Research Council research fellowship on 'Risk measurement in financial institutions' (RES-000-27-0014). Cotter's contribution to the study has been supported by a University College Dublin School of Business research grant. The authors would like to thank two reviewers and the editor for their comments.

# 1. INTRODUCTION

Since they arose in the early 1990s, risk managers have come to rely heavily on models that forecast the risks associated with financial portfolios. Often known as Value-at-Risk (VaR) models, these models in fact can and sometimes do forecast the complete density functions of prospective financial losses (or, equivalently, financial returns). The outputs of these models can then be used to forecast a variety of different measures of financial risk: these include measures such as the VaR and the Expected Shortfall (ES), but also families of risk measures such as coherent, spectral and distortion risk measures.[1] However, these risk forecasts are inevitably open to error – the density functions might be misspecified (giving rise to model risk), and model parameters are unknown, which forces risk managers to rely on estimated parameters and exposes them to parameter risk – and it is therefore important that risk managers have some idea of the precision or accuracy of the forecasts on which they are relying.

This paper investigates this issue, and addresses three particular questions related to the precision of risk forecasts:

- How can we estimate the precision of different risk forecasts? This is not a new question, and there is considerable literature on it (see section 3 below). However, as we shall see, existing approaches are limited in a number of ways, and the present paper proposes a more flexible Monte-Carlo approach that is free of many of the limitations of the approaches proposed so far.

- What do we know of the distributions of estimators of financial risk measures? In fact, we know from existing statistical theory that these distributions are asymptotically normal, and practitioners often rely on such asymptotic results as practical short-cuts. Unfortunately, we do not know how large the sample sizes must be for asymptotic results to be taken seriously. Thus, investigating the finite sample properties of these estimators is of considerable practical importance.

- What can we say about the relationship between the precision of risk forecasts and the underlying loss (or return) density function? So, for example, can we say anything about how estimates of precision might be affected by factors such as skewness or tail heaviness in the underlying loss distribution? This is a very difficult question to answer in a general way, but we suggest a procedure that provides some useful insights into these issues and into related questions such as the relative precision of estimators of different financial risk measures.

---

[1] Dowd (2005, ch. 2) has an overview of these different risk measures and their properties.

This paper is organized as follows. Section 2 discusses the risk measures to be considered: the VaR, the Expected Shortfall, and the Spectral Risk Measures (SRMs).[2] Section 3 discusses the existing literature on the precision of estimators of financial risk measures, and section 4 discusses alternative estimators of precision. Section 5 sets out our methodology for evaluating the precision of our risk-measure estimators, and section 6 looks into the difficult issue of the relationship between the precision of these estimators and the underlying loss (or return) distribution. Section 7 concludes.

## 2. ALTERNATIVE RISK MEASURES

Suppose our underlying random variable is the realised daily loss (which is positive for an actual loss, and negative for a profit) on a portfolio. If the confidence level is $\alpha$, our first risk measure is the VaR at this confidence level, i.e.:

$$VaR_\alpha = q_\alpha \qquad (1)$$

where $q_\alpha$ is the $\alpha$-quantile of the loss distribution. The VaR is the most widely used financial risk measure, but has been heavily criticized in recent years for some of its properties (e.g., its lack of subadditivity; see, e.g., Artzner *et al.*, 1999). Note, therefore, that the VaR is defined in terms of a conditioning parameter, the confidence level, the value of which usually needs to be specified – more or less arbitrarily – by the user.

Our second risk measure is the ES, which can be defined as the average of the worst $1-\alpha$ of losses. In the case of a continuous loss distribution, the ES is given by:

---

[2] The risk measures considered include those most commonly used by financial risk measures. Distortion risk measures have a different epistemological foundation and are widely used in actuarial circles. They are closely related to coherent risk measures and many risk measures – such as the ES – are members of both the coherent and distortion families.

$$ES_\alpha = \frac{1}{1-\alpha}\int_\alpha^1 q_p \, dp \tag{2}$$

The ES gives equal weight to each of the worst $1-\alpha$ of losses and no weight to any other observations. The ES is superior to the VaR in a number of respects (e.g., it is subadditive and coherent). However, the ES is specified in terms of the same conditioning parameter as the VaR and, as with the VaR, there is often little to tell us what value this parameter should take.

Our third measure is a Spectral Risk Measure (SRM). Following Acerbi (2002), consider a risk measure $M_\phi$ defined by:

$$M_\phi = \int_0^1 q_p \phi(p) \, dp \tag{3}$$

where $\phi(p)$ is a weighting function defined over $p$, the cumulative probabilities in the range between 0 and 1. Borrowing from Acerbi (2004, proposition 3.4), the risk measure $M_\phi$ is coherent if and only if $\phi(p)$ satisfies the following properties:

- *Positivity*: $\phi(p) \geq 0$, i.e., weights are always non-negative.
- *Normalisation*: $\int_0^1 \phi(p) \, dp = 1$, i.e., weights sum to one.
- *Increasingness*: $\phi'(p) \geq 0$, i.e., higher losses have weights that are higher than or equal to those of smaller losses.[3]

We now need to specify a suitable weighting function $\phi(p)$, and a good choice is the following exponential function:

$$\phi(p) = \frac{ke^{-k(1-p)}}{1-e^{-k}} \tag{4}$$

---

[3] Strictly speaking, Acerbi's proposition refers to a more general class of risk measures, whereas we are concerned only with those measures that Acerbi defines as non-singular risk measures (i.e., we are not concerned with the singular risk measures that he mentions and then dismisses as uninteresting).

where the coefficient *k* is the user's degree of absolute risk aversion. The function $\phi(p)$ can also be interpreted as the user's risk-aversion function. The user's risk aversion means that higher losses attract higher weights than small losses, and the more risk-averse the user, the more rapidly the weights will rise. The risk measure itself then can then be obtained by substituting (4) into (3), viz.:

$$M_\phi = \int_0^1 \frac{ke^{-k(1-p)}}{1-e^{-k}} q_p dp = \frac{ke^{-k}}{1-e^{-k}} \int_0^1 e^{kp} q_p dp \qquad (5)$$

Thus, an SRM has the attractive property that it takes account of the user's risk aversion. Furthermore, an SRM based on an exponential risk-aversion function is predicated on a single conditioning factor, the user's degree of absolute risk aversion. And, unlike the earlier conditioning parameter, the value of this parameter is unique (i.e., because it is determined by the user's risk-aversion).

## 3. EXISTING LITERATURE ON THE PRECISION OF RISK ESTIMATORS

Naturally, we never actually know the values of our risk measures in practice, because the parameters of the loss distribution will be unknown. (This is true even in the favourable unlikely case where the form of the distribution itself is known, but we will ignore this issue here.) We must therefore work with estimates of these parameters, and this means that we must deal with *estimators* of our risk measures. (We will henceforth call these "risk estimators" for short, to avoid the correct but ungainly term "risk-measure estimators".) This then raises a key question: how can we evaluate the precision – or more loosely, the accuracy – of risk estimators?

Before we begin to answer this question ourselves, we should first consider the answers provided in established literature, and the principal findings of 22 studies in this literature are summarised in Table 1.[4] This shows that existing studies differ enormously in how they have addressed the precision issue. It also shows that many of these studies are limited in one way or another:

- The majority of them only apply to one risk measure, and typically the VaR.

---

[4] We would emphasize too that the selection of studies shown in this Table is at best a good illustration of the existing literature, and is by no means selective. The Table also omits studies that have looked at precision from a Bayesian perspective (e.g., Dowd, 2000; Siu *et ali,* 2001).

- Some approaches are limited to a single distribution (e.g., Jorion (1996) and Chappell and Dowd (1999) require that losses be normal).
- A number of approaches only give estimates of standard errors (i.e., and do not give confidence intervals). However, as discussed in the next section, standard errors can give misleading impressions of the precision of risk estimators.
- A considerable number of approaches are based on asymptotic theory, and asymptotic results might not be appropriate with the sample sizes that practitioners often have to work with.[5]

**Insert Table 1 here**

Some studies also report results on the relative precision of different risk estimators. For example, several studies find that VaR and ES estimators have comparable precision when loss distributions are normal or close to normal, but ES estimators decline in precision relative to VaR estimators as tails become heavier (e.g., Yamai and Yoshiba (2002), Acerbi (2004)). However, such findings are essentially illustrative and one of the purposes of the present study is to shed more light on these and related issues.

## 4. PRECISION ESTIMATORS AND THE DISTRIBUTION OF RISK ESTIMATORS

We also we need to consider how to estimate precision itself. One obvious way to do so is to estimate the standard error (SE) of a risk estimator. The SE is helpful in its own right as a (rough and ready) estimate of precision, and can also be used to construct confidence intervals using textbook formulas. However, as Yamai and Yoshiba (2002) note, comparisons of the SEs across a set of different risk measures are complicated by the fact that the different risk measures considered typically have different values. For example, the SE for risk measure $A$ might be larger than the SE for risk measure $B$, so in such circumstances it does always make sense to say that risk measure *A* is estimated less precisely than risk measure *B*? The answer is obviously 'no': the first risk measure might have a much bigger value than the second, and its SE might be only marginally bigger than that of the second risk measure. In this case, the 'absolute' SE for *A* would be larger than that of *B*, but in relative terms – that is, relative to the value of the risk measure itself – *A* is estimated more precisely than *B*. We should therefore estimate precision taking account of the estimated value of the risk measure, and can do so by working with a 'standardized' SE rather than an 'absolute' SE, i.e., we work with the SE divided by a point estimate of the relevant risk measure.[6]

---

[5] Nor is this list of limitations exhaustive. The majority of them also apply to unconditional risk estimators (although some exceptions are McNeil and Frey (2000), Giannopoulos and Tunaru (2004) and Chen and Tang (2005)). Exactly how conditional dependence would affect precision is, as yet, a difficult question to answer.

A second precision estimator is a confidence interval. A confidence interval is often more helpful than an SE because it can give a more 'complete' picture of precision: it can take more account of the distribution of risk estimators and so allow for factors such as asymmetries, heavy-tails, etc., that can cause the SE to give a 'misleading' impression of precision. However, as with SEs, comparison of 'raw' confidence intervals can be problematic when the risk measures themselves have different values, so we work here with confidence intervals in standardized form, i.e., with 'raw' confidence intervals divided by point estimates of risk measures.

It is also useful to examine the related issue of the distribution of the risk estimators themselves. In fact, it is well-known in the statistical literature that linear combinations of order statistics are asymptotically normally distributed (see, e.g., Stiegler (1974), Mason (1981)), and this result implies that estimators of all three of our risk measures should be asymptotically normal. However, knowing that the distribution of risk estimators approaches normality in the limit as our sample size gets large does *not* tell us whether it is safe to assume that they are normally distributed for any given (finite) sample size: we don't know how quickly the estimators converge to normality as the sample size increases. Hence, if we are thinking of using asymptotic results, it is prudent to check if the distribution of risk estimators is 'close enough' to normality to allow those results to apply.

Such information is also important in determining the usefulness of several alternative precision estimators:

---

[6] However, standardization is not without its drawbacks as we can get meaningless results when the risk estimator goes to zero.

- The SE as a precision indicator implicitly presupposes that the distribution is symmetric, and where distributions are asymmetric, we really need an alternative indicator that allows for this asymmetry.

- The SE is often used as an input to textbook formulas for confidence intervals, but this practice is only defensible if the underlying risk estimators are suitably 'well-behaved (e.g., symmetric, and normally or *t* distributed), and this 'well-behavedness' condition might not hold empirically. Where such assumptions do not hold, we should use properly constructed confidence intervals (e.g., Monte Carlo-based intervals) instead.

## 5. METHODOLOGY

We wish now to get some sense of how the precision of risk estimators varies across the different types of risk measure. However, this task is complicated by: (a) the variety of precision estimators available; (b) the possibility that results will depend on sample size and/or the conditioning parameter; and (c) the likelihood that results will depend on the underlying loss distribution (e.g., results might depend on the skewness or kurtosis of this distribution). Handling (a) and (b) is relatively straightforward, but dealing with (c) is more difficult because we cannot search over every plausible loss distribution. We therefore need a meaningful way to restrict our search, whilst also attempting to draw out conclusions of (hopefully) more general validity.

      We also need some way of organising the search. One reasonable approach is based on the principle that we start from a simple benchmark distribution and get some sense of the precision of the different risk estimators in this benchmark case. A simple and well-understood (and therefore natural) benchmark is a standard normal distribution. We then need some way of generalising from this benchmark case to see how generalisation might affect our results. We would suggest thinking of generalisation in terms of the successive relaxation of the moment restrictions implied by our chosen benchmark (i.e., that the mean should be 0, the variance should be 1, the skewness should be 0 and the kurtosis should be 3). We initially generalise by relaxing the first and second moment (i.e., mean and variance) restrictions, and so move from a standard normal to an unrestricted normal. After this, we generalise further by successively relaxing the third and fourth moment (i.e., skewness and

kurtosis) restrictions. Naturally, we recognise that there are different ways of relaxing these moment restrictions, and there is no uniquely 'best' way to do: all we can do here is suggest a plausible procedure and then regard any conclusions we draw from this analysis as tentative hypotheses that would be subject to confirmation (or rejection) by later studies.

The precise approach is as follows. We first select a range of sample sizes, and in this paper we choose *n* equal to 250, 500, 1000 and 2000. These values correspond to sample sizes of 1, 2, 4 and 8 trading years at 250 trading days to a year: this is a good range, because risk practitioners would not usually work with sample sizes that are less than 1 trading year or more than 8 trading years. In fact, many practitioners would work with sample sizes of 250 or 500, so it is the shorter end of the sample range that we should mainly be interested in. We then select some illustrative parameter values for our risk measures (i.e., we choose values for the confidence level for our VaR and ES risk measures, and values for the coefficient of absolute risk aversion in the case of our SRM). We chose fairly standard confidence-level values of 90%, 95% and 99%, and we chose a fairly wide range of ARA values equal to 5, 25 and 100. Having made these calibrations, we then implement the following three-stage procedure.

- *Stage One*: We select a standard normal benchmark and assume that losses are standard normal. We then estimate the precision of our different risk estimators using our two different precision estimators across the range of selected sample sizes, and thence evaluate how the precision of our risk estimators in the standard normal case changes as we alter the way that precision is measured and as we vary the sample size.

- *Stage Two*: We then consider how our results might change in the face of changes in the mean $\mu$ and standard deviation $\sigma$ of portfolio losses. More specifically, we compare the cases of $(\mu=5, \sigma=1)$ and $(\mu=0, \sigma=5)$ against our standard normal benchmark $(\mu=0, \sigma=1)$. The former gives an example of a non-standard mean, and the latter an example of a non-standard variance. Given that the normal is 'well-behaved' and well understood, these two alternative cases should suffice to give a fairly complete picture of the sensitivities of results to changes in $\mu$ and $\sigma$.

- *Stage Three*: Given that the normal restricts the skewness and kurtosis to 0 and 3 respectively, we now select two convenient cases involving skewness and excess kurtosis. More particularly, the impact of skewness is examined by selecting a two-part normal (2PN) distribution, where the latter is calibrated to produce a skewness of about 0.492.[7] Comparing this 2PN against the standard normal gives us a comparison between a distribution that has no skew and one that has a fairly pronounced skew, and this range of skewnesses encompasses those commonly observed in financial returns. The impact of excess kurtosis is captured by a zero-mean, unit-standard deviation $t$ distribution with 5 degrees of freedom.[8] This distribution has a kurtosis of 9, which is higher than the kurtoses usually reported for financial returns. Thus our comparison of the $t$ and the standard normal implies that we are comparing a range of kurtoses from 3 to 9, and this range includes the kurtoses usually found for financial returns.

All calculations were carried out using parametric Monte Carlo simulation with 10000 simulation trials in each case: we run 10000 trials under the specified loss distribution, and estimate the various risk measures from the 'sample' order statistics generated in each trial;[9] we then estimate the (standardised) SE and (standardised) confidence interval from the 10000 sets of 'sample' risk estimates obtained in this way.

---

[7] The 2PN can be represented in various ways, but the particular distribution chosen here is the ($\mu, \sigma_1, \sigma_2$) representation of the 2PN with $\mu = 0$, $\sigma_1 = 1.3$ and $\sigma_2 = 0.65$. These parameter values ensure that our 2PN has zero mean, unit variance, a skewness of about 0.492 and a 'small' (and hopefully negligible) excess kurtosis (equal to 0.148). For more on this distribution, see John (1982).

[8] This distribution has a pdf equal to $\sqrt{(v-2)/v}$ times the pdf of a Student-$t$ with $v$ degrees of freedom (taken in our calibrations to be 5). This distribution has zero mean, unit variance, zero skew, and (provided $v > 4$, which is necessary for the kurtosis to exist) a kurtosis equal to $3(v-2)/(v-4)$. For more on the Student-$t$, see Evans *et alia* (2000, p.180).

[9] To be more precise: if we want the quantile or VaR at the, say, 90% confidence level, and we have 1000 loss observations in our simulated sample, we follow the usual practice and take this quantile/VaR estimate to be the 101$^{st}$ highest loss observation; and we then take the ES estimate as the average of the 100 highest loss observations. More generally, for a confidence level α and sample size $n$, we take the quantile/VaR at the α confidence level to be equal to the (1-α)$n$+1$^{st}$ sample order statistic. The SRM estimate is then taken as the suitably weighted average of our VaR/quantile estimates.

## 6. RESULTS

**Stage One: Standard Normal Losses**

We begin by examining the distributions and precision measures of our risk estimators under standard normality.

*VaR results*

The results reported in Table 2 suggest that the standard normal VaR estimators have the following properties:

- Their means rise with $\alpha$ and are invariant to *n*, as expected.
- Their standard errors rise with $\alpha$ and fall with *n*, as expected.
- Their skewnesses tend to be positive, rise with $\alpha$, fall with *n*, and go to zero as *n* gets large.
- Their kurtoses tend to exceed the normal kurtosis (i.e., 3), rise with $\alpha$, fall with *n*, and approach 3 as *n* gets large.
- Their Jarque-Bera (JB) test results are not supportive of normality, except where $\alpha$ is low and *n* high.
- The precision results indicate no clear pattern as $\alpha$ rises (which is perhaps a little surprising given that we might have expected that precision would consistently fall with the high values of $\alpha$ that we are considering), but they do indicate that precision rises with *n* (which *is* what we would expect).

**Insert Table 2 here**

The third, fourth and fifth findings suggest that VaR estimators tend to normality as *n* gets large, but they also approach normality more slowly as $\alpha$ gets larger. This impression is confirmed by Figure 1, which shows histograms for the 90% and 99% VaRs for sample sizes of 250 and 500. This Figure indicates that the 99% VaR estimators are notably non-normal, especially for the smaller sample size: they also indicate that the convergence of 99% VaR estimators to normality is slow. This suggests that, *with the sample sizes often available, practitioners would be unwise to assume that VaR estimators are normally or even close to normally* *distributed, i.e., in particular, they should be wary of using results based on asymptotic normality theory.*

**Insert Figure 1 here**

This conclusion also has two other useful corrolaries. (1) If the distribution of VaR estimators is not symmetric, then practitioners should not use the SE as a precision measure. (2) If the distribution of VaR estimators is not normal, then practitioners should be careful about estimating VaR confidence intervals by inserting estimates of SEs into textbook formulas for confidence intervals, because those formulas might not apply.

*ES results*

The ES results reported in Table 3 and illustrated in Figure 2 are similar to the VaR ones in most ways (e.g. they generally exhibit a positive skewness, have comparable precision, etc.) and the only other noteworthy features are the following:

- The skewness, kurtosis and JB results usually suggest that ES estimators are a little 'closer to normal' than the earlier VaR estimators.
- The precision measures are now 'well-behaved' in the sense that they indicate that precision falls with $\alpha$ as well as rises with *n*. Taken together, these first two bullet points suggest that *ES estimators are a little 'better behaved' than VaR estimators*.
- Most importantly, in this standard normal case, *the precision of ES estimators is of much the same order of magnitude as that of VaR estimators.*

**Insert Table 3 here**
**Insert Figure 2 here**

*SRM results*

The corresponding results for the SRM risk measure are shown in Table 4 and illustrated in Figure 3. These suggest that the ARA coefficient plays much the same role in SRMs as the $\alpha$ parameter plays with the VaR and ES. These results are broadly similar to the earlier ones, and it is particularly interesting to note that *estimators of all three risk measures have similar precision*. These results also suggest that SRM estimators tend to be a little bit closer to normal than the ES estimators, and are certainly closer to normal than the VaR estimators.

**Insert Table 4 here**
**Insert Figure 3 here**

**Stage Two: Non-Standard Normal Losses (Impact of Mean and Variance)**
Having established results for the benchmark case where losses are standard normal, we now investigate how results might change as we allow for changes in the mean and standard deviation. To do so, we compare results based on three hypothetical sets of parameter values: $\mu = 0$ and $\sigma = 1$ (our benchmark case); $\mu = 5$ and $\sigma = 1$; and $\mu = 0$ and $\sigma = 5$. A comparison of the first and second cases allows us to investigate the impact of a change in $\mu$; and a comparison of the first and third cases allows us to investigate the impact of a change in $\sigma$.

Our results are clear. For all risk measures, changes in the mean and/or standard deviation have no impact on the higher moments of the distribution of risk estimators (and therefore have no impact on the skewness, kurtosis and JB test results). Furthermore, changes in the mean and standard deviation have the impacts we might expect on a priori grounds. More specifically, we get the following results, given in Table 5 to 7:

An increase in $\mu$:
- impacts the means of the risk estimators *pari passu* (in the cases of VaR and ES) or close to *pari passu* (in the case of the SRMs);
- has a 'small' negative impact on the standard error, which declines as *n* gets larger; and
- has a 'moderate' widening impact on the confidence intervals, and this impact declines as *n* gets larger.

An increase in $\sigma$:

- leads to major increases in the means of the risk estimators, and these increases are of broadly the same order of magnitude across the different risk measures and are greater for higher $\alpha$; and
- leads to the same precision estimates as in the standard normal case. Thus precision estimators are 'well behaved' and are of broadly the same magnitude across the risk measures.

Drawing these findings together, perhaps the most significant conclusion is that *for normally distributed losses, estimates of precision are of much the same order of magnitude across the different types of risk estimator*.[10]

**Insert Table 5 here**
**Insert Table 6 here**
**Insert Table 7 here**

**Stage Three: Non-Normal Losses (Impact of Skewness and Kurtosis)**

*Impact of skewness*

The skewness results are presented in Tables 8. These are presented as the relevant 2PN estimate divided by the corresponding standard normal estimate. This format makes it easy to see the impact that skewness makes. These results paint a very clear picture, i.e., introducing skewness:

- has a notably positive impact on the standard errors;
- has a negligible effect on the width of the confidence intervals;

and these results hold for all risk estimators.

**Insert Table 8 here**

*Impact of kurtosis*

Tables 9 give the corresponding kurtosis results, in this case expressed as the ratio of the precision estimates generated under our specified *t* distribution divided by the

---

[10] These findings should be no surprise given that the normal is so well-understood. Nor should it be any surprise that some of these results have also been established analytically: for example, McNeil *et alia* (2005, pp. 39, 45) provide analytical results showing how the unrestricted normal VaR and ES are related to their standard normal counterparts.

corresponding estimators generated under the standard normal. The main highlights are:

- Risk estimators under the *t*-distribution are always less precise than their counterparts under standard normality, and this suggests that *excess kurtosis makes risk estimators less precise.*

- Increasing the conditioning parameter (i.e., depending on the risk measure, the confidence level or the degree of risk aversion) makes risk estimators under the heavy-tailed distribution less precise, relative to their counterparts under standard normality.

- By and large, the ratios are somewhat higher for the ES and SRM estimators. This suggests that *tail heaviness has a greater (though not much greater) deleterious effect on the precision of ES and SRM estimators than on the precision of VaR estimators.*

**Insert Table 9 here**

## 7. CONCLUSIONS

This paper addresses three main issues. The first is the question, how can we estimate the precision of different risk estimates? Various methods have been suggested in the existing literature, but many existing methods are subject to significant limitations: they apply to one risk measure only (typically the VaR), or are limited to specific distributions (e.g., the normal distribution), or only give estimates of standard errors (i.e., and don't give estimates of confidence intervals), or rely on asymptotic theory (which may not be appropriate empirically). We suggest an approach based on Monte Carlo simulation that is free of the above limitations.[11]

The second issue addressed is the distribution of risk estimators. We know from existing statistical theory that the distribution of risk estimators is asymptotically normal. However, this theory does not tell us how quickly estimators converge to normality, and the results presented here indicate that this convergence is sufficiently slow that practitioners working with 1-day forecast horizons will often not have samples long enough for them to invoke asymptotic normality. Thus, *for most practical purposes we cannot rely on risk estimators to be normally distributed with the sample sizes often available.*[12]

---

[11] As hinted at in footnote 5, a limitation of our approach is that we restrict our attention to unconditional estimators and ignore time dependence in losses or returns. We believe that most of our findings would also apply to conditional estimators that take account of time dependence (see the next note), but investigating the impact of such extensions would be an involved task beyond the scope of our present study. Another limitation worth pointing out is that we do not consider how estimators might be improved using variance-reduction methods. For on this latter issue, see Inui and Kijima (2004).

The final issue addressed is the question of how the precision of risk estimators might be affected by the underlying loss distribution. This is a very difficult question to answer in a general way, but we suggest a procedure that highlights the moments of the loss distribution: we start with a standard normal which restricts what each of the first four moments should be; we then relax each of these moment restrictions in turn and see what effect the relaxation has on our precision estimates. This procedure generates some insightful results, including the following:

- When the loss distribution is normal, estimators of *all three risk measures have similar precision*.
- The impact of skewness on precision depends on how we measure precision: introducing skewness has a noticeable positive impact on (standardized) standard errors, but no notable impact on (correctly estimated standardized) confidence intervals.
- Introducing excess kurtosis into the loss distribution has the effect of *making all risk estimators less precise than they were*, and it *reduces the precision of ES and SRM estimators somewhat more than it reduces the precision of VaR estimators*.

Of course, we recognise that these results were obtained for specified distributions, and it is possible that a different set of distributions might lead to somewhat different conclusions. We therefore offer these conclusions as tentative hypotheses – albeit plausible hypotheses – that other researchers might wish to investigate further.

---

[12] These results were generated under the assumption of unconditional normal losses. However, statistical intuition would suggest that our results about the distribution of risk estimators are likely to be robust: for example, if losses are heavier tailed than the normal, or conditionally distributed, then we would often expect these changes to slow down the convergence to normality even further. Thus, if anything, we would suggest that our results about the slowness of convergence to asymptotic normality are likely to be over-optimistic, and this would reinforce our warning about the dangers of invoking asymptotic results in a practical risk measurement context.

# FIGURES

# FIGURE 1: DISTRIBUTIONS OF STANDARD NORMAL VAR ESTIMATORS

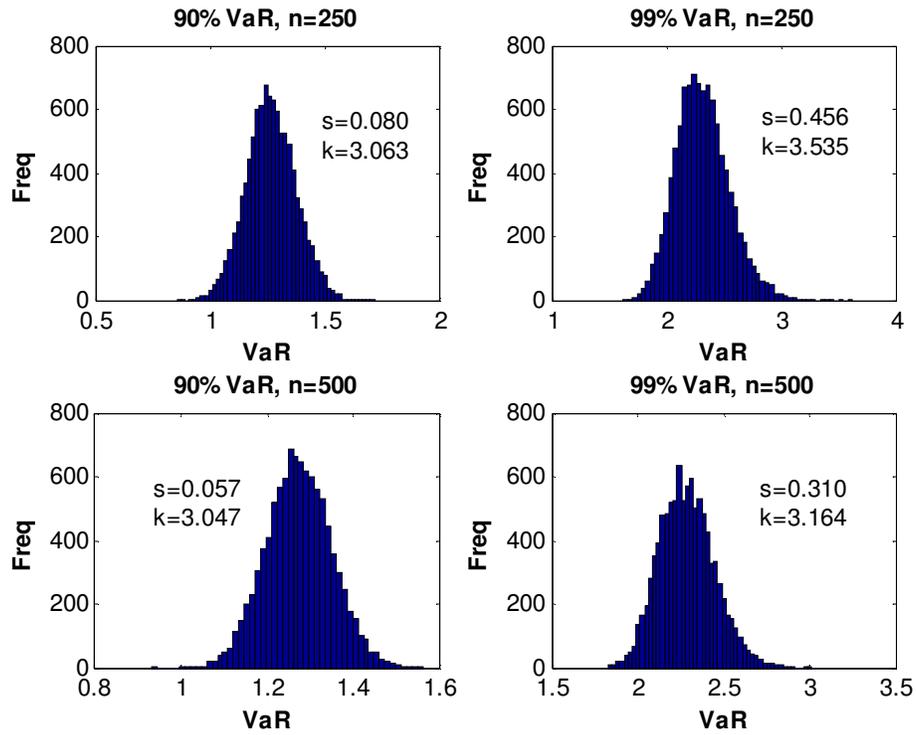

Notes: Based on 10000 iid Monte Carlo simulations using a N(0,1) loss distribution, with the VaR estimators taken as the relevant order statistic from each trial sample. *s* and *k* are the estimated coefficients of skewness and kurtosis.

**FIGURE 2: DISTRIBUTIONS OF STANDARD NORMAL ES ESTIMATORS**

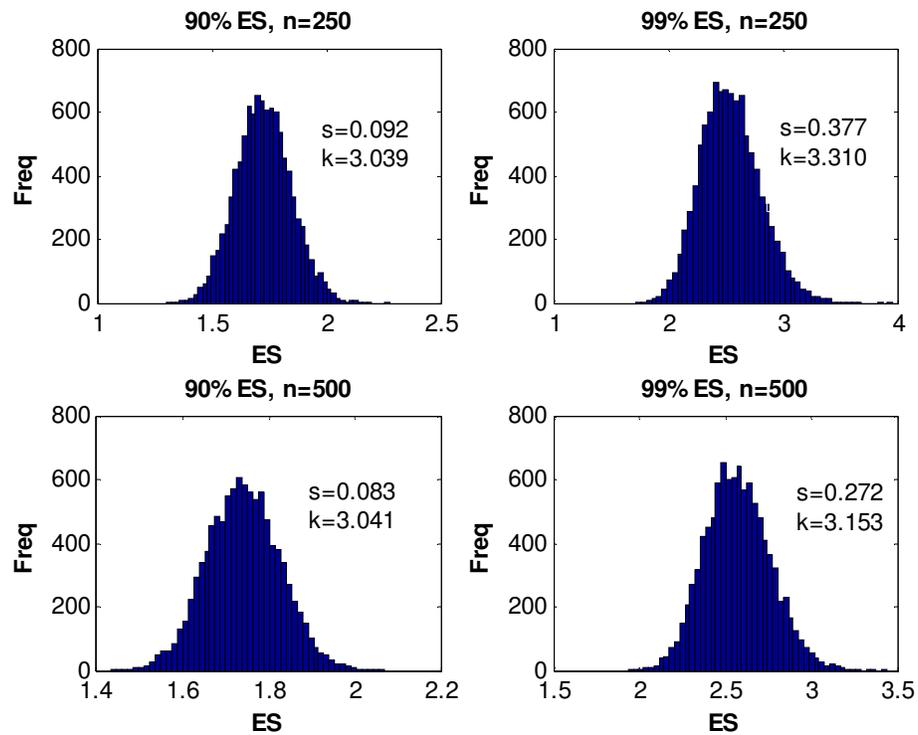

Notes: Based on 10000 iid Monte Carlo simulations using a N(0,1) loss distribution. *s* and *k* are the estimated coefficients of skewness and kurtosis.

# FIGURE 3: DISTRIBUTIONS OF STANDARD NORMAL SRM ESTIMATORS

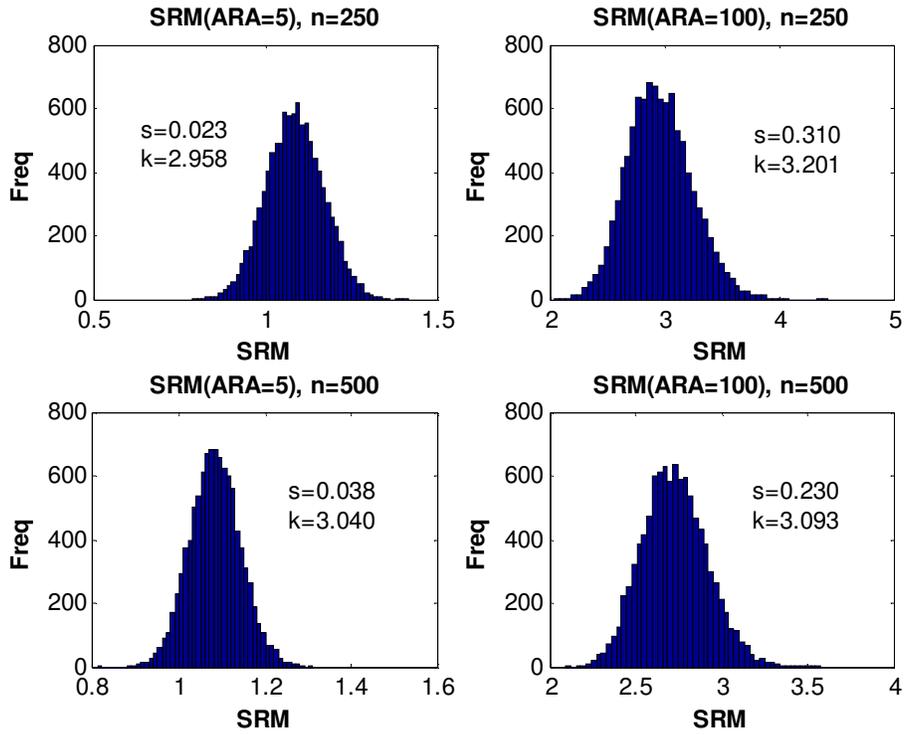

Notes: Based on 10000 iid Monte Carlo simulations using a N(0,1) loss distribution. *s* and *k* are the estimated coefficients of skewness and kurtosis.

# TABLES

## TABLE 1: STUDIES THAT HAVE ADDRESSED THE PRECISION OF RISK ESTIMATORS

| *Study* | *Relevant Findings* |
|---|---|
| Kendall and Stuart (1972) | Derives formula for asymptotic variance of quantile estimator: for quantile $x$ at confidence level $\alpha$ and density $f(x)$, this variance is $\alpha(1-\alpha)/(n[f(x)]^2)$. |
| Reiss (1976) | Deals with asymptotic expansions for variance of sample quantiles; more accurate than Kendall-Stuart formula, but not so tractable. |
| Jorion (1996) | Obtains standard error formula for quantile where losses are normally distributed. |
| Pritsker (1997) | Examines precision of VaR estimators using Monte-Carlo simulation. |
| Butler and Schachter (1998) | Proposes kernel and bootstrap methods to estimate the precision of VaR estimators. |
| Chappell and Dowd (1999) | Uses variance-ratio theory to obtain confidence intervals for normal VaR. |
| Gourieroux et alia (2000) | Show asymptotic normality of kernel estimators of VaR. |
| McNeil and Frey (2000) | Uses profile maximum likelihood to estimate confidence intervals for VaR. |
| Dowd (2001) | Uses order-statistics theory to obtain confidence intervals for VaR. |
| Mausser (2001) | Uses *L*-estimators to improve precision of VaR estimators. |
| Yamai and Yoshiba (2002) | Obtains formulas for asymptotic standard deviations of VaR and ES estimators. Provides simulation results for stable Paretian distributions suggesting that VaR and ES estimators have comparable precision for moderately sized tails, but ES estimator becomes much less precise relative to VaR estimators as tails become heavy. |
| Acerbi (2004) | Obtains asymptotic variances of estimators of VaR, ES and SRM. Provides simulation results for lognormal and power-law distributions suggesting that VaR and ES estimators have comparable precision for moderately sized tails, but ES estimator declines in precision relative to VaR estimators as tails become heavy. |
| Giannopoulos and Tunaru (2004) | Applies filtered historical simulation to obtain standard errors of VaR and ES. Empirical results suggest that ES estimators are considerably less precise than VaR estimators. |
| Inui and Kijima (2004) | Proposes extrapolation method to increase the accuracy of VaR and ES estimators. Presents results for *t* distributions showing that ordinary VaR estimators can be strongly biased, and this bias increases as tails become heavier. |
| Scaillet (2004) | Shows asymptotic normality of kernel estimators of ES. |
| Chen (2005) | Suggests an improved kernel method for the estimation of ES. |
| Chen and Tang (2005) | Examines standard errors of nonparametric VaR estimators for dependent financial returns. |
| Dowd (2005) | Extends Dowd (2001) to obtain confidence intervals for ES. |
| Manistre and Hancock (2005) | Derives asymptotic variance of ES estimator, and suggests that this has good finite-sample properties. Discusses how ES estimators can be made more accurate using variance-reduction methods. |
| McNeil et alia (2005) | Provides some analytical results for VaR and ES. |
| Cotter and Dowd (2006) | Applies a parametric bootstrap approach to estimate extreme risks for VaR, ES and SRMs for equity futures data. Results suggest that ES is estimated relatively more precisely than VaR, but that SRM estimators are notably less precise than estimators of VaR or ES. |
| Gourieroux and Liu (2006) | Provides general formula for asymptotic distribution of nonparametric estimator of distortion risk measures. Provides a number of formulas for variances of VaR and ES estimators for special-case distributions. |

**TABLE 2: RESULTS FOR STANDARD NORMAL VALUE AT RISK**

**(a) MOMENT RESULTS**

| $\alpha$ | n=250 | n=500 | n=1000 | n=2000 |
|---|---|---|---|---|
| **Means** | | | | |
| 0.9 | 1.267 | 1.2739 | 1.278 | 1.2802 |
| 0.95 | 1.6409 | 1.6333 | 1.639 | 1.6425 |
| 0.99 | 2.3169 | 2.2867 | 2.3063 | 2.3161 |
| **Standard deviations** | | | | |
| 0.9 | 0.1086 | 0.0752 | 0.0544 | 0.0385 |
| 0.95 | 0.134 | 0.0927 | 0.066 | 0.0469 |
| 0.99 | 0.2324 | 0.1624 | 0.1151 | 0.0823 |
| **Skewnesses** | | | | |
| 0.9 | 0.0779 | 0.0897 | 0.06 | 0.0182 |
| 0.95 | 0.132 | 0.0922 | 0.0948 | -0.004 |
| 0.99 | 0.4563 | 0.2774 | 0.1857 | 0.1461 |
| **Kurtoses** | | | | |
| 0.9 | 3.0633 | 3.0227 | 3.0891 | 2.9468 |
| 0.95 | 3.1111 | 3.0523 | 3.0583 | 3.1139 |
| 0.99 | 3.5354 | 3.2659 | 3.0265 | 3.0587 |
| **Jarque-Bera prob-values** | | | | |
| 0.9 | 0.0028 | 0.0011 | 0.0095 | 0.42 |
| 0.95 | 0 | 0.0005 | 0.0003 | 0.0662 |
| 0.99 | 0 | 0 | 0 | 0 |

**(b) PRECISION MEASURES**

**Standardised standard errors**

| $\alpha$ | n=250 | n=500 | n=1000 | n=2000 |
|---|---|---|---|---|
| 0.9 | 0.0857 | 0.0591 | 0.0426 | 0.0301 |
| 0.95 | 0.0817 | 0.0568 | 0.0403 | 0.0285 |
| 0.99 | 0.1003 | 0.071 | 0.0499 | 0.0356 |

**Bounds of standardised 90% confidence intervals**

| $\alpha$ | n=250 LB | n=250 UB | n=500 LB | n=500 UB | n=1000 LB | n=1000 UB | n=2000 LB | n=2000 UB |
|---|---|---|---|---|---|---|---|---|
| 0.9 | 0.8605 | 1.1447 | 0.9048 | 1.0991 | 0.9306 | 1.0707 | 0.9506 | 1.0494 |
| 0.95 | 0.8706 | 1.1367 | 0.9079 | 1.0964 | 0.9342 | 1.0679 | 0.9529 | 1.0466 |
| 0.99 | 0.8481 | 1.1747 | 0.8881 | 1.1236 | 0.921 | 1.0852 | 0.943 | 1.0601 |

Notes: Based on 10000 Monte Carlo simulation trials. $\alpha$ is the confidence level, $n$ is the sample size, and LB and UB refer to the lower and upper bounds of the confidence intervals.

**TABLE 3: RESULTS FOR STANDARD NORMAL EXPECTED SHORTFALL**

**(a) MOMENT RESULTS**

| $\alpha$ | n=250 | n=500 | n=1000 | n=2000 |
|---|---|---|---|---|
| **Means** | | | | |
| 0.9 | 1.726 | 1.74 | 1.748 | 1.7516 |
| 0.95 | 2.0294 | 2.0373 | 2.0509 | 2.0565 |
| 0.99 | 2.5457 | 2.5727 | 2.6185 | 2.6399 |
| **Standard deviations** | | | | |
| 0.9 | 0.1203 | 0.0853 | 0.0603 | 0.0431 |
| 0.95 | 0.1532 | 0.1095 | 0.0769 | 0.055 |
| 0.99 | 0.2621 | 0.1907 | 0.1381 | 0.1002 |
| **Skewnesses** | | | | |
| 0.9 | 0.0915 | 0.0475 | 0.0636 | -0.0266 |
| 0.95 | 0.1566 | 0.0758 | 0.0927 | -0.0022 |
| 0.99 | 0.3772 | 0.2183 | 0.1836 | 0.1664 |
| **Kurtoses** | | | | |
| 0.9 | 3.0388 | 2.9504 | 3.0795 | 3.0334 |
| 0.95 | 3.0759 | 2.9818 | 3.0621 | 3.0599 |
| 0.99 | 3.3069 | 3.1645 | 3.0254 | 3.1241 |
| **Jarque-Bera prob-values** | | | | |
| 0.9 | 0.0007 | 0.0915 | 0.0092 | 0.4401 |
| 0.95 | 0 | 0.0077 | 0.0003 | 0.4721 |
| 0.99 | 0 | 0 | 0 | 0 |

**(b) PRECISION MEASURES**

**Standardised standard errors**

| $\alpha$ | n=250 | n=500 | n=1000 | n=2000 |
|---|---|---|---|---|
| 0.9 | 0.0697 | 0.049 | 0.0345 | 0.0246 |
| 0.95 | 0.0755 | 0.0537 | 0.0375 | 0.0267 |
| 0.99 | 0.103 | 0.0741 | 0.0528 | 0.038 |

**Bounds of standardised 90% confidence intervals**

| $\alpha$ | n=250 LB | n=250 UB | n=500 LB | n=500 UB | n=1000 LB | n=1000 UB | n=2000 LB | n=2000 UB |
|---|---|---|---|---|---|---|---|---|
| 0.9 | 0.8865 | 1.116 | 0.9208 | 1.0821 | 0.9438 | 1.0576 | 0.9594 | 1.0403 |
| 0.95 | 0.8799 | 1.1267 | 0.9123 | 1.0892 | 0.9392 | 1.0633 | 0.9564 | 1.0441 |
| 0.99 | 0.8438 | 1.1762 | 0.8815 | 1.1271 | 0.9175 | 1.09 | 0.9394 | 1.0633 |

Notes: Based on 10000 Monte Carlo simulation trials. $\alpha$ is the confidence level, $n$ is the sample size, and LB and UB refer to the lower and upper bounds of the confidence intervals.

**TABLE 4: RESULTS FOR STANDARD NORMAL SPECTRAL RISK MEASURE**

**(a) MOMENT RESULTS**

| ARA | n=250 | n=500 | n=1000 | n=2000 |
|---|---|---|---|---|
| **Means** | | | | |
| 5 | 1.0863 | 1.0837 | 1.0831 | 1.0825 |
| 25 | 2.0326 | 1.9931 | 1.975 | 1.9647 |
| 100 | 2.9583 | 2.7264 | 2.6166 | 2.5597 |
| **Standard deviations** | | | | |
| 5 | 0.0836 | 0.0582 | 0.0411 | 0.0292 |
| 25 | 0.149 | 0.104 | 0.0719 | 0.0509 |
| 100 | 0.2823 | 0.1885 | 0.1264 | 0.088 |
| **Skewnesses** | | | | |
| 5 | 0.0226 | 0.0308 | 0.0379 | -0.0338 |
| 25 | 0.1378 | 0.0577 | 0.0755 | -0.0027 |
| 100 | 0.3099 | 0.1776 | 0.1573 | 0.1205 |
| **Kurtoses** | | | | |
| 5 | 2.958 | 2.9743 | 3.0832 | 3.0537 |
| 25 | 3.0654 | 2.9686 | 3.0575 | 3.0422 |
| 100 | 3.2012 | 3.1162 | 3.0388 | 3.0973 |
| **Jarque-Bera prob-values** | | | | |
| 5 | 0.4526 | 0.3946 | 0.0716 | 0.2113 |
| 25 | 0 | 0.051 | 0.0044 | 0.6854 |
| 100 | 0 | 0 | 0 | 0 |

**(b) PRECISION MEASURES**

**Standardised standard errors**

| ARA | n=250 | n=500 | n=1000 | n=2000 |
|---|---|---|---|---|
| 5 | 0.077 | 0.0537 | 0.0379 | 0.027 |
| 25 | 0.0733 | 0.0522 | 0.0364 | 0.0259 |
| 100 | 0.0954 | 0.0691 | 0.0483 | 0.0344 |

**Bounds of standardised 90% confidence intervals**

| ARA | n=250 LB | n=250 UB | n=500 LB | n=500 UB | n=1000 LB | n=1000 UB | n=2000 LB | n=2000 UB |
|---|---|---|---|---|---|---|---|---|
| 5 | 0.874 | 1.1263 | 0.9123 | 1.0901 | 0.9378 | 1.0633 | 0.9552 | 1.0443 |
| 25 | 0.8824 | 1.1243 | 0.9153 | 1.0869 | 0.9406 | 1.0609 | 0.9571 | 1.043 |
| 100 | 0.853 | 1.1628 | 0.8889 | 1.1182 | 0.9239 | 1.0821 | 0.9442 | 1.0577 |

Notes: Based on 10000 Monte Carlo simulation trials. ARA is the coefficient of absolute risk aversion, $n$ is the sample size, and LB and UB refer to the lower and upper bounds of the confidence intervals.

## TABLE 5: RESULTS FOR NORMAL VALUE AT RISK

**Standardised standard errors**

| $\alpha$ | n=250 | n=500 | n=1000 | n=2000 |
|---|---|---|---|---|
| **N(0,1)** | | | | |
| 0.9 | 0.0857 | 0.0591 | 0.0426 | 0.0301 |
| 0.95 | 0.0817 | 0.0568 | 0.0403 | 0.0285 |
| 0.99 | 0.1003 | 0.071 | 0.0499 | 0.0356 |
| **N(5,1)** | | | | |
| 0.9 | 0.0173 | 0.012 | 0.0087 | 0.0061 |
| 0.95 | 0.0202 | 0.014 | 0.0099 | 0.0071 |
| 0.99 | 0.0318 | 0.0223 | 0.0158 | 0.0113 |
| **N(0,5)** | | | | |
| 0.9 | 0.0857 | 0.0591 | 0.0426 | 0.0301 |
| 0.95 | 0.0817 | 0.0568 | 0.0403 | 0.0285 |
| 0.99 | 0.1003 | 0.071 | 0.0499 | 0.0356 |

**Bounds of standardised 90% confidence intervals**

| $\alpha$ | n=250 LB | n=250 UB | n=500 LB | n=500 UB | n=1000 LB | n=1000 UB | n=2000 LB | n=2000 UB |
|---|---|---|---|---|---|---|---|---|
| **N(0,1)** | | | | | | | | |
| 0.9 | 0.8605 | 1.1447 | 0.9048 | 1.0991 | 0.9306 | 1.0707 | 0.9506 | 1.0494 |
| 0.95 | 0.8706 | 1.1367 | 0.9079 | 1.0964 | 0.9342 | 1.0679 | 0.9529 | 1.0466 |
| 0.99 | 0.8481 | 1.1747 | 0.8881 | 1.1236 | 0.921 | 1.0852 | 0.943 | 1.0601 |
| **N(5,1)** | | | | | | | | |
| 0.9 | 0.9718 | 1.0293 | 0.9807 | 1.0201 | 0.9859 | 1.0144 | 0.9899 | 1.0101 |
| 0.95 | 0.968 | 1.0338 | 0.9773 | 1.0237 | 0.9838 | 1.0168 | 0.9884 | 1.0115 |
| 0.99 | 0.9519 | 1.0553 | 0.9649 | 1.0388 | 0.9751 | 1.0269 | 0.9819 | 1.019 |
| **N(0,5)** | | | | | | | | |
| 0.9 | 0.8605 | 1.1447 | 0.9048 | 1.0991 | 0.9306 | 1.0707 | 0.9506 | 1.0494 |
| 0.95 | 0.8706 | 1.1367 | 0.9079 | 1.0964 | 0.9342 | 1.0679 | 0.9529 | 1.0466 |
| 0.99 | 0.8481 | 1.1747 | 0.8881 | 1.1236 | 0.921 | 1.0852 | 0.943 | 1.0601 |

Notes: Based on 10000 Monte Carlo simulation trials. $\alpha$ is the confidence level, *n* is the sample size, and LB and UB refer to the lower and upper bounds of the confidence intervals.

## TABLE 6: RESULTS FOR NORMAL EXPECTED SHORTFALL

| | | **Standardized standard errors** | | | |
|---|---|---|---|---|---|
| | $\alpha$ | n=250 | N=500 | n=1000 | n=2000 |
| | | **N(0,1)** | | | |
| | 0.9 | 0.0697 | 0.049 | 0.0345 | 0.0246 |
| | 0.95 | 0.0755 | 0.0537 | 0.0375 | 0.0267 |
| | 0.99 | 0.103 | 0.0741 | 0.0528 | 0.038 |
| | | **N(5,1)** | | | |
| | 0.9 | 0.0179 | 0.0127 | 0.0089 | 0.0064 |
| | 0.95 | 0.0218 | 0.0156 | 0.0109 | 0.0078 |
| | 0.99 | 0.0347 | 0.0252 | 0.0181 | 0.0131 |
| | | **N(0,5)** | | | |
| | 0.9 | 0.0697 | 0.049 | 0.0345 | 0.0246 |
| | 0.95 | 0.0755 | 0.0537 | 0.0375 | 0.0267 |
| | 0.99 | 0.103 | 0.0741 | 0.0528 | 0.038 |

| | | | **Bounds of standardised 90% confidence intervals** | | | | |
|---|---|---|---|---|---|---|---|
| $\alpha$ | n=250 | n=250 | n=500 | N=500 | n=1000 | n=1000 | n=2000 | n=2000 |
| | LB | UB | LB | UB | LB | UB | LB | UB |
| | | | | **N(0,1)** | | | | |
| 0.9 | 0.8865 | 1.116 | 0.9208 | 1.0821 | 0.9438 | 1.0576 | 0.9594 | 1.0403 |
| 0.95 | 0.8799 | 1.1267 | 0.9123 | 1.0892 | 0.9392 | 1.0633 | 0.9564 | 1.0441 |
| 0.99 | 0.8438 | 1.1762 | 0.8815 | 1.1271 | 0.9175 | 1.09 | 0.9394 | 1.0633 |
| | | | | **N(5,1)** | | | | |
| 0.9 | 0.9709 | 1.0298 | 0.9796 | 1.0212 | 0.9855 | 1.0149 | 0.9895 | 1.0105 |
| 0.95 | 0.9653 | 1.0366 | 0.9746 | 1.0258 | 0.9823 | 1.0184 | 0.9873 | 1.0128 |
| 0.99 | 0.9473 | 1.0594 | 0.9598 | 1.0432 | 0.9716 | 1.0309 | 0.9791 | 1.0219 |
| | | | | **N(0,5)** | | | | |
| 0.9 | 0.8865 | 1.116 | 0.9208 | 1.0821 | 0.9438 | 1.0576 | 0.9594 | 1.0403 |
| 0.95 | 0.8799 | 1.1267 | 0.9123 | 1.0892 | 0.9392 | 1.0633 | 0.9564 | 1.0441 |
| 0.99 | 0.8438 | 1.1762 | 0.8815 | 1.1271 | 0.9175 | 1.09 | 0.9394 | 1.0633 |

Notes: Based on 10000 Monte Carlo simulation trials. $\alpha$ is the confidence level, $n$ is the sample size, and LB and UB refer to the lower and upper bounds of the confidence intervals.

**TABLE 7: RESULTS FOR NORMAL SPECTRAL RISK MEASURE**

| | | **Standardised standard errors** | | | |
|---|---|---|---|---|---|
| | ARA | n=250 | n=500 | n=1000 | n=2000 |
| | | **N(0,1)** | | | |
| | 5 | 0.077 | 0.0537 | 0.0379 | 0.027 |
| | 25 | 0.0733 | 0.0522 | 0.0364 | 0.0259 |
| | 100 | 0.0954 | 0.0691 | 0.0483 | 0.0344 |
| | | **N(5,1)** | | | |
| | 5 | 0.0136 | 0.0095 | 0.0067 | 0.0048 |
| | 25 | 0.0205 | 0.0146 | 0.0102 | 0.0073 |
| | 100 | 0.0313 | 0.0229 | 0.0161 | 0.0115 |
| | | **N(0,5)** | | | |
| | 5 | 0.077 | 0.0537 | 0.0379 | 0.027 |
| | 25 | 0.0733 | 0.0522 | 0.0364 | 0.0259 |
| | 100 | 0.0954 | 0.0691 | 0.0483 | 0.0344 |

| | **Bounds of standardised 90% confidence intervals** | | | | | | | |
|---|---|---|---|---|---|---|---|---|
| ARA | n=250 | n=250 | n=500 | n=500 | n=1000 | n=1000 | n=2000 | n=2000 |
| 0 | LB | UB | LB | UB | LB | UB | LB | UB |
| | | | **N(0,1)** | | | | | |
| 5 | 0.874 | 1.1263 | 0.9123 | 1.0901 | 0.9378 | 1.0633 | 0.9552 | 1.0443 |
| 25 | 0.8824 | 1.1243 | 0.9153 | 1.0869 | 0.9406 | 1.0609 | 0.9571 | 1.043 |
| 100 | 0.853 | 1.1628 | 0.8889 | 1.1182 | 0.9239 | 1.0821 | 0.9442 | 1.0577 |
| | | | **N(5,1)** | | | | | |
| 5 | 0.9777 | 1.0224 | 0.9844 | 1.016 | 0.9889 | 1.0112 | 0.992 | 1.0079 |
| 25 | 0.9672 | 1.0347 | 0.9763 | 1.0243 | 0.9833 | 1.0171 | 0.988 | 1.0121 |
| 100 | 0.9518 | 1.0534 | 0.9632 | 1.0391 | 0.9747 | 1.0273 | 0.9814 | 1.0192 |
| | | | **N(0,5)** | | | | | |
| 5 | 0.874 | 1.1263 | 0.9123 | 1.0901 | 0.9378 | 1.0633 | 0.9552 | 1.0443 |
| 25 | 0.8824 | 1.1243 | 0.9153 | 1.0869 | 0.9406 | 1.0609 | 0.9571 | 1.043 |
| 100 | 0.853 | 1.1628 | 0.8889 | 1.1182 | 0.9239 | 1.0821 | 0.9442 | 1.0577 |

Notes: Based on 10000 Monte Carlo simulation trials. ARA is the coefficient of absolute risk aversion, $n$ is the sample size, and LB and UB refer to the lower and upper bounds of the confidence intervals.

**TABLE 8: RATIOS OF STATISTICS UNDER 2PN DISTRIBUTION TO THOSE UNDER STANDARD NORMAL DISTRIBUTION**

**VAR RISK MEASURE**

**Standardised standard errors**

| $\alpha$ | n=250 | n=500 | n=1000 | n=2000 |
|---|---|---|---|---|
| 0.9 | 1.130 | 1.162 | 1.134 | 1.140 |
| 0.95 | 1.120 | 1.134 | 1.127 | 1.137 |
| 0.99 | 1.106 | 1.087 | 1.092 | 1.110 |

**Bounds of standardised 90% confidence intervals**

| $\alpha$ | n=250 LB | n=250 UB | n=500 LB | n=500 UB | n=1000 LB | n=1000 UB | n=2000 LB | n=2000 UB |
|---|---|---|---|---|---|---|---|---|
| 0.9 | 0.981 | 1.012 | 0.986 | 1.014 | 0.988 | 1.011 | 0.997 | 1.009 |
| 0.95 | 0.980 | 1.015 | 0.988 | 1.008 | 0.992 | 1.009 | 0.995 | 1.007 |
| 0.99 | 0.981 | 1.019 | 0.991 | 1.006 | 0.991 | 1.007 | 0.995 | 1.004 |

**ES RISK MEASURE**

**Standardized standard errors**

| $\alpha$ | n=250 | n=500 | n=1000 | n=2000 |
|---|---|---|---|---|
| 0.9 | 1.118 | 1.118 | 1.125 | 1.114 |
| 0.95 | 1.106 | 1.104 | 1.117 | 1.112 |
| 0.99 | 1.100 | 1.096 | 1.100 | 1.100 |

**Bounds of standardized 90% confidence intervals**

| $\alpha$ | n=250 LB | n=250 UB | n=500 LB | n=500 UB | n=1000 LB | n=1000 UB | n=2000 LB | n=2000 UB |
|---|---|---|---|---|---|---|---|---|
| 0.9 | 0.986 | 1.012 | 0.988 | 1.008 | 0.992 | 1.006 | 0.995 | 1.004 |
| 0.95 | 0.984 | 1.011 | 0.992 | 1.009 | 0.992 | 1.006 | 0.996 | 1.005 |
| 0.99 | 0.978 | 1.019 | 0.990 | 1.010 | 0.989 | 1.008 | 0.993 | 1.006 |

**SRM RISK MEASURE**

**Standardised standard errors**

| ARA | n=250 | n=500 | n=1000 | n=2000 |
|---|---|---|---|---|
| 5 | 1.110 | 1.140 | 1.137 | 1.126 |
| 25 | 1.112 | 1.113 | 1.121 | 1.112 |
| 100 | 1.112 | 1.097 | 1.112 | 1.105 |

**Bounds of standardised 90% confidence intervals**

| ARA | n=250 LB | n=250 UB | n=500 LB | n=500 UB | n=1000 LB | n=1000 UB | n=2000 LB | n=2000 UB |
|---|---|---|---|---|---|---|---|---|
| 5 | 0.984 | 1.013 | 0.988 | 1.010 | 0.991 | 1.007 | 0.995 | 1.006 |
| 25 | 0.986 | 1.012 | 0.989 | 1.009 | 0.992 | 1.008 | 0.996 | 1.005 |
| 100 | 0.980 | 1.016 | 0.989 | 1.009 | 0.989 | 1.008 | 0.994 | 1.005 |

Notes: Based on 10000 Monte Carlo simulation trials. The Table reports the ratios of the relevant statistic for a 2PN distribution (with mean 0, std 1, and $\sigma_1 = 1.35$; this has a skewness of 0.492 and a 'small' excess kurtosis of 0.148) to the relevant statistic for a standard normal distribution. $\alpha$ is the confidence level, $n$ is the sample size, and LB and UB refer to the lower and upper bounds of the confidence intervals.

# TABLE 9: RATIOS OF STATISTICS UNDER t DISTRIBUTION TO THOSE UNDER STANDARD NORMAL DISTRIBUTION

### VAR RISK MEASURE

**Standardised standard errors**

| $\alpha$ | n=250 | n=500 | n=1000 | n=2000 |
|---|---|---|---|---|
| 0.9 | 1.168 | 1.203 | 1.195 | 1.169 |
| 0.95 | 1.332 | 1.333 | 1.340 | 1.323 |
| 0.99 | 1.756 | 1.672 | 1.695 | 1.666 |

**Bounds of standardised 90% confidence intervals**

| $\alpha$ | n=250 LB | n=250 UB | n=500 LB | n=500 UB | n=1000 LB | n=1000 UB | n=2000 LB | n=2000 UB |
|---|---|---|---|---|---|---|---|---|
| 0.9 | 0.978 | 1.025 | 0.979 | 1.018 | 0.986 | 1.014 | 0.992 | 1.009 |
| 0.95 | 0.957 | 1.049 | 0.970 | 1.029 | 0.979 | 1.023 | 0.986 | 1.016 |
| 0.99 | 0.897 | 1.131 | 0.928 | 1.078 | 0.947 | 1.059 | 0.964 | 1.040 |

### ES RISK MEASURE

**Standardised standard errors**

| $\alpha$ | n=250 | n=500 | n=1000 | n=2000 |
|---|---|---|---|---|
| 0.9 | 1.515 | 1.533 | 1.522 | 1.537 |
| 0.95 | 1.722 | 1.698 | 1.715 | 1.730 |
| 0.99 | 2.199 | 2.157 | 2.186 | 2.232 |

**Bounds of standardised 90% confidence intervals**

| $\alpha$ | n=250 LB | n=250 UB | n=500 LB | n=500 UB | n=1000 LB | n=1000 UB | n=2000 LB | n=2000 UB |
|---|---|---|---|---|---|---|---|---|
| 0.9 | 0.948 | 1.062 | 0.958 | 1.044 | 0.972 | 1.030 | 0.981 | 1.022 |
| 0.95 | 0.925 | 1.094 | 0.944 | 1.064 | 0.959 | 1.043 | 0.971 | 1.034 |
| 0.99 | 0.852 | 1.201 | 0.886 | 1.146 | 0.910 | 1.101 | 0.932 | 1.081 |

### SRM RISK MEASURE

**Standardised standard errors**

| ARA | n=250 | n=500 | n=1000 | n=2000 |
|---|---|---|---|---|
| 5 | 1.344 | 1.348 | 1.335 | 1.333 |
| 25 | 1.850 | 1.831 | 1.832 | 1.853 |
| 100 | 2.273 | 2.285 | 2.302 | 2.369 |

**Bounds of standardised 90% confidence intervals**

| ARA | n=250 LB | n=250 UB | n=500 LB | n=500 UB | n=1000 LB | n=1000 UB | n=2000 LB | n=2000 UB |
|---|---|---|---|---|---|---|---|---|
| 5 | 0.962 | 1.050 | 0.969 | 1.029 | 0.979 | 1.021 | 0.986 | 1.015 |
| 25 | 0.915 | 1.105 | 0.935 | 1.073 | 0.954 | 1.051 | 0.968 | 1.037 |
| 100 | 0.858 | 1.200 | 0.891 | 1.144 | 0.913 | 1.106 | 0.935 | 1.080 |

Notes: Based on 10000 Monte Carlo simulation trials. The Table reports the ratios of the relevant statistic for a t distribution (with mean 0, std 1, and 5 degrees of freedom) to the relevant statistic for a standard normal distribution. $\alpha$ is the confidence level, *n* is the sample size, and LB and UB refer to the lower and upper bounds of the confidence intervals.